\documentclass[aps,prl,twocolumn,showpacs,preprintnumbers,amsmath,amssymb]{revtex4}

\usepackage{graphicx}
\usepackage{dcolumn}
\usepackage{bm}

\begin{document}


\title{Self-reptation and slow topological time scale of  
knotted polymers}

\author{Enzo Orlandini$^{1,2}$}
\author{Attilio L. Stella$^{1,2}$}%
\author{Carlo Vanderzande$^{3,4}$}
\author{Francesco Zonta$^5$}
\affiliation{%
$^1$Dipartimento di Fisica and CNR-INFM,  Universit\`{a} di Padova, I-35131, Padova, Italy.\\ 
$^2$ Sezione INFN, Universit\`{a} di Padova, I-35131 Padova, Italy.\\
$^3$Departement WNI, Hasselt University, 3590 Diepenbeek, Belgium. \\
$^4$Instituut Theoretische Fysica, Katholieke Universiteit Leuven, 3001 Heverlee, Belgium.\\
$^5$Dipartimento di Fisica, Universit\`{a} di Padova, I-35131, Padova, Italy.
} 


\date{\today}

\begin{abstract}
We investigate the Rouse dynamics of a flexible ring polymer with a prime knot. Within a Monte Carlo
approach, we locate the knot, follow its diffusion, and observe the 
fluctuations of its length. We characterise a topological time scale, and 
show that it is related to a self-reptation of the knotted region.  The associated dynamical exponent, $z_T=2.32\pm.1$,  can be related to that of the equilibrium knot length distribution and determines the behaviour of several dynamical quantities.
\end{abstract}

\pacs{36.20.Ey, 64.60.Ht, 02.10.Kn, 87.15.Aa}
\maketitle

In the physics of polymers, mutual and self entanglements play an essential role \cite{DE86}. A 
particularly relevant type of entanglement is 
associated with the presence of knots. Indeed, it has been known 
for some time that long ring polymers inevitably contain a knot \cite{D62SW88}. Topology is also of much 
interest for biopolymers, where knots have been found in the DNA of viruses and bacteria \cite{DNAT05}, 
and also in some proteins \cite{KnotProt}. 
The recent experimental possibility to tie knots in DNA double strands or actin filaments \cite{Arai99}, 
and to study their diffusion in the presence of a stretching force \cite{ExpQ03}, has further increased the 
interest in topology related issues among polymer physicists. 

So far, the statistical physics of knotted polymers has focused mostly on static properties \cite{SEo}. For 
example, it was found that the presence of a knot does not alter the exponent $\nu$ that relates the radius 
of gyration $R_G(N)$ to the length $N$ of the polymer, $R_G(N) \sim N^\nu$,  but only influences scaling 
corrections \cite{OTBS98,BEAF05}. There have been fewer studies of dynamical scaling properties of knotted
polymers, even though the associated time scales could be relevant to describe gel electrophoresis,
folding of knotted proteins or other dynamical processes. 
Simulations have given evidence that a peculiar, topological, characteristic time $\tau_{T}$ determines the
decay of the radius of gyration autocorrelation function of such polymers \cite{Quake94,PYL02}. This decay
time appears longer than that observed for open or closed unknotted chains.
As a function of chain length, it was found to scale with a dynamical exponent 
$z_{T}$, that could not be determined precisely but whose value probably is bigger than the Rouse one. So far, the physical origin of this time was not understood.

In this Letter, we investigate 
how a knot influences dynamical scaling properties of flexible polymers in a good solvent within the 
framework of Rouse dynamics \cite{DE86,Rouse53}. In a simulation, we directly follow the motion of the knotted region and the fluctuations of its length (measured along the backbone). We develop a simple picture, based on reptation, that establishes a link between $z_{T}$ and the exponent governing the distribution of this length.

In Rouse dynamics, one models a polymer as a set of $N+1$ beads (monomers) that are connected by harmonic 
springs, and that are subjected to random thermal forces exerted by the solvent. The motion of 
the monomers is described by a Langevin equation, which for ideal chains can be solved using a 
transformation to normal coordinates. If self-avoidance is taken into account, Rouse dynamics can no longer 
be solved exactly, but scaling arguments together with numerical results \cite{dGB79}, show that the 
center of mass of the polymer, $\vec{R}_{cm}$, performs ordinary diffusion, i.e.
\begin{eqnarray}
g_3(t,N)\equiv\langle \left(\vec{R}_{cm}(t)-\vec{R}_{cm}(0)\right)^2\rangle \simeq 6D_{cm}t
\label{1}
\end{eqnarray}
where the average is taken over realisations of the stochastic process. The diffusion constant is inversely 
proportional to the length $N$ of the polymer, i.e. $D_{cm} \sim N^{-1}$. 
The autocorrelation function of the end-to-end vector decays exponentially with a time constant $\tau_R$ that 
grows as
\begin{eqnarray}
\tau_R \sim N^{2\nu+1}
\label{2}
\end{eqnarray}
with $N$. In $d=3$, $2\nu+1\simeq2.2$ \cite{dGB79}. We will refer to $\tau_R$ as the {\it Rouse time scale}. 
It can  be interpreted as the time that the center of mass of the polymer needs to diffuse 
over a length equal to its 
radius of gyration, $\tau_R \sim R_G^2(N)/D_{cm} \sim N^{2\nu+1}$.

The diffusion of one particular monomer, or of a segment of $m+1$ monomers, is described by the function 
\begin{eqnarray}
g_1(t,N) =\frac{1}{m+1} \sum_{i=(N-m)/2}^{(N+m)/2} \langle\left(\vec{R}_i(t)-\vec{R}_i(0)\right)^2\rangle.
\label{3}
\end{eqnarray}
It is known \cite{KB84GK86} that $g_1(t)$ has a scaling form 
\begin{eqnarray}
g_1(t,N) \simeq t^\beta F(t/\tau_R).
\label{4}
\end{eqnarray}
The function $F$ describes the crossover between an initial power law regime ($F(x) \to $ constant, for $x \to 0$) and a late time regime for 
which a monomer has to follow the diffusion of the whole polymer as given by (\ref{1}). This implies a power 
law form for $F$ at late times and $\beta=2\nu/(2\nu+1)$, i.e. in the initial regime the movement of a 
segment of the polymer is subdiffusive. Notice that $g_1(\tau_R,N) \sim R_G^2(N)$, so that the Rouse 
time scale also corresponds to the time for one monomer to diffuse over a distance equal to $R_G(N)$. 
In the whole dynamics, it is the only relevant time.

While Rouse dynamics neglects important physical effects, such as hydrodynamic interactions between 
the monomers, several experimental situations are known by now for which it gives an adequate description.
As an example, we mention that  current fluorescence techniques allow to follow the motion of 
individual monomers in, e.g.,  DNA-chains and hence to directly determine a function such as 
$g_1(t,N)$. Measurements of this type  on double stranded DNA have recently seen diffusive 
behaviour of individual monomers consistent with Rouse behaviour \cite{SAGK04}.

A crucial quantity to characterise the dynamics of a knotted polymer turns out to be the length of the knot.  
A precise definition for this quantity can be given for flat 
knots \cite{MHDKK02}. These are knots in a polymer that is strongly adsorbed to a plane or 
constrained between two walls. In this context, it was found that the length $l_k$ of a knot $k$  is a fluctuating 
quantity, whose equilibrium distribution 
function is a power law, $p(l_k) \sim l_k^{-c} G(l_k/N)$ where $G$ is a scaling function.
It follows, that the average length of the knot scales with $N$ as  
$\langle l_k \rangle \sim N^\sigma$, with $\sigma=\max[0,2-c]$. In a good solvent, flat knots were found to be 
strongly localised ($\sigma=0$) \cite{MHDKK02}, while below the $\theta$-transition, they 
delocalise ($\sigma=1$) \cite{EAC03}.  In
order to extend these results to 
three dimensions, 
one needs a good definition of the length of a knot. This should correspond to the intuitive idea that it measures
 the portion of the
polymer backbone which ``hosts'' the knot entanglement \cite{KOVDS00}. 
Recently, a powerful computational approach to  determine such a length, and its scaling properties, was introduced~\cite{BEAF05}. In this 
method, for a given knotted ring polymer, various open portions are considered and for each of these a closure is made by joining its ends with 
an off-lattice path. The length of the knot in a given configuration can then be identified with 
the shortest portion still displaying the original knot. 
In this way, it was found that in good solvent, three dimensional knots are {\it weakly localised} 
with an exponent $\sigma  = .75\pm .05$. 

To simulate the dynamics of knotted polymers, we start from a simple self-avoiding polygon (SAP) on a cubic lattice with a knot in it. Most of our calculations were done 
with a trefoil ($3_1$) knot. After performing a number of BFACF \cite{BFACF} moves to relax the configuration, the 
resulting SAP evolves according to a $N$-conserving Monte-Carlo dynamics with local moves only. This is 
expected to give a proper description of Rouse dynamics \cite{SimRouse}. During the subsequent evolution, using the computational techniques developed in \cite{BEAF05},
observables such as the length of the knot, the radius of gyration of the whole polymer, the location 
of the center of mass of the polygon and the location of the center of mass of the knot, $\vec{R}_{cm,k}$, 
are computed. 
We have performed calculations for various $N \leq 400$. 

\begin{figure}[here]
\includegraphics[width=7.0cm]{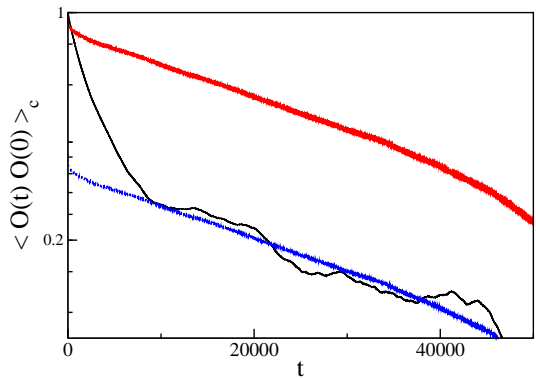}
\caption{\label{fig3} (Color online) 
Semilogarithmic plot of the autocorrelation functions versus time 
for a SAP with $N=400$. 
Shown are results for the radius of gyration (black, below) and the knot length (red, above). The blue line was obtained by shifting the red one vertically.}
\end{figure}

\begin{figure}[here]
\includegraphics[width=7.0cm]{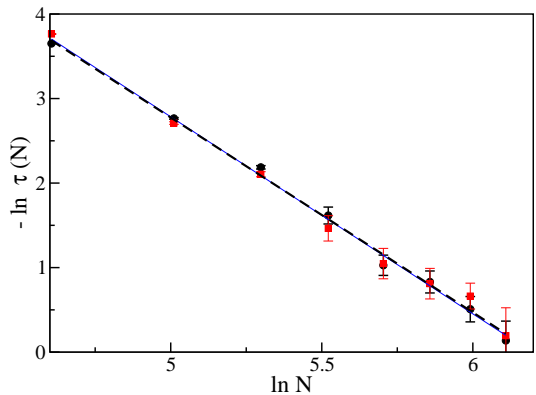}
\caption{\label{fig4} (Color online) 
Log-log plot of the topological time scale versus $N$ as determined 
from the radius of gyration autocorrelation function (black circles) 
and from the knot length autocorrelation (red squares). 
The fitted lines have slopes $-2.31 \pm 0.08$ (black dashed line) and $-2.33 \pm 0.08$  (blue full line)
respectively.}
\end{figure}

From the data on the knot length as a function of time, we can construct 
the knot length autocorrelation function  $\langle l_k(t)l_k(0)\rangle_c$ (where the subscript 'c' indicates  the properly normalised connected autocorrelation). In Fig. \ref{fig3}, the top (red) line shows our results for this quantity for $N=400$.  This autocorrelation function has a simple exponential decay. Similar behaviour is found for other $N$ values. Fig. \ref{fig4} presents our data for the decay time constants as a function of $N$.
The behaviour is power law, and the exponent is $2.33 \pm 0.08$. The value of this exponent is higher than that of Rouse dynamics. Further evidence that this presents a new time scale comes from an investigation of the autocorrelation function of the radius of gyration of the polymer.
In Fig. \ref{fig3} we also plotted our results for this quantity (bottom line).
Clearly, it has a double exponential decay.
In comparison, for an unknot, we observe a pure exponential decay.  A careful analysis shows that the time constant of the unknot, 
together with the fast one of the knotted polymer, 
are proportional to $\tau_R$. On the other hand, as can be seen in Fig. \ref{fig3},
the late time decay of the radius of gyration autocorrelation occurs with the same time constant as that of the length autocorrelation. A quantitative analysis of its $N$-dependence supports this conclusion (see Fig. \ref{fig4}). Indeed, the value of the associated exponent equals 
$2.31\pm 0.08$, consistent within the numerical accuracy with that determined from the length autocorrelation.
Together, these data therefore provide strong evidence for the presence of a new, slow time scale $\tau_{T} \sim N^{z_{T}}$, with $z_{T}=2.32\pm.1$,  in the dynamics of knotted polymers. The data on the knot length autocorrelation show that this scale corresponds to the time over which the knot length decorrelates.

To check if the results are influenced by the topology of the knot, we have performed a completely similar 
study for the figure-eight ($4_1$) knot. Again the radius of gyration 
autocorrelation function decays as a double exponential, with a fast time scale that can be identified 
with $\tau_R$. The slower time scale grows with $N$ with an exponent whose numerical value,
$2.33\pm.08$ is consistent with that found for the trefoil. Moreover, the same exponent was found to govern the decay of the $4_1$ length 
autocorrelation function.

A simple argument explaining the value of $z_{T}$ is based on a reptation picture. Reptation describes the 
motion of a polymer in a dense environment of other polymers that restrict its movement to a tube \cite{dGRep81,DE86}. This 
is, for example, the case in a polymer melt. 
Through a diffusive motion along the tube axis, first one end and later the whole polymer moves out of the original tube, creating in this way a new one. This process is slow and it takes a time $\tau_d \sim N^3$ for the new and the old tube to decorrelate. In the case of interest here, we can imagine that the knot, which is weakly localized, creates a local tube that constrains the movement of the monomers it contains. Since it is only the knotted part, of length proportional to $N^\sigma$,  that performs the reptation in this case, we expect a decorrelation of the knot tube after a time $N^{3 \sigma}$. Taking the estimate of $\sigma$ from \cite{BEAF05}, we find $3\sigma=2.25 \pm .15$, which 
is consistent with the identification $z_{T}=3\sigma$. This agreement supports the plausibility of the self-reptation mechanism.

In order to determine whether the topological time also influences other dynamical properties, we have investigated the diffusion of the center of mass of the whole polymer and of the knotted region. From our data, we determine $g_3(t,N)$ and a similar quantity for the  
knotted part of the polymer
\begin{eqnarray}
g_{3,k}(t,N)=\langle \left(\vec{R}_{cm,k}(t)-\vec{R}_{cm,k}(0)\right)^2\rangle 
\label{5}
\end{eqnarray}
This function has some similarity to $g_1(t,N)$ since it describes the motion of a segment of the polymer. 
Hence we can expect it to have a scaling behaviour similar to (\ref{4}). There are however differences since the 
number of monomers in the knot increases with $N^\sigma$ on average, and, at fixed $N$, fluctuates in time.

\begin{figure}[here]
\includegraphics[width=7.0cm]{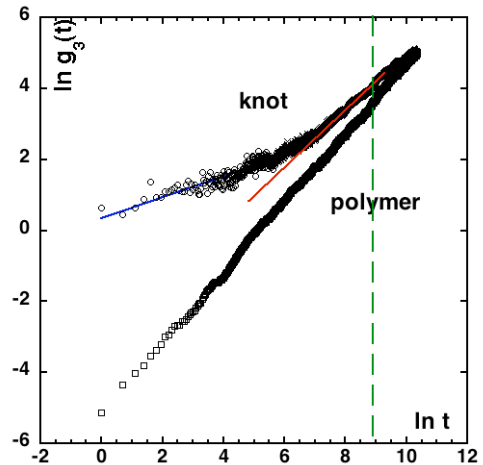}
\caption{\label{fig1} (Color online) The function $g_3$ of the whole polymer (lower curve) and of the knot (upper curve) ($N=250$). The straight lines are best fits through the initial (blue) and intermediate (red) time regime of the knot. Their slopes are $.29$ and $.78$, respectively. The dashed vertical line correspond to $\tau_R$. }
\end{figure}

In Fig. \ref{fig1}, we show a typical result ($N=250$) for the functions $g_3(t,N)$ and $g_{3,k}(t,N)$. 
Our data for the diffusion of the center of mass of the whole polymer are fully consistent with Eq.
(\ref{1}) and with the relation $D_{cm} \sim N^{-1}$, a strong indication that the diffusion of the 
polymer as a whole is not changed by the presence of the knot \cite{PYL02}. 

\begin{figure}[here]
\includegraphics[width=7.0cm]{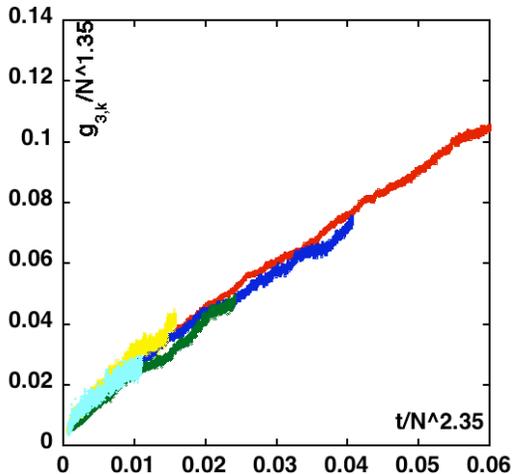}
\caption{\label{fig2} (Color online) 
Scaling plot of our results for $g_{3,k}/N^{z_{T}\gamma}$ 
versus $t/N^{z_{T}}$ for various $N$-values using 
$z_{T}=2.35, \gamma=.57$. Different colors correspond to different
$N$ values,
red: $N=$150, blue: 200, dark green: 250, yellow: 300 and light green: 350. }
\end{figure}

More interesting is the diffusion of the knot itself. 
As can be seen in Fig. \ref{fig1}, $g_{3,k}$ shows several distinct power law regimes. The value of 
the estimated exponent for the early time region fluctuates with $N$, but a good average is $0.27\pm 0.1$. This initial regime ends after a 
time $\tau_1$, which grows as a power of $N$, $\tau_1 \sim N^{z_1}$, where $z_1=1.97\pm.1$. After a short 
crossover, in an intermediate time regime, the behaviour is again power law, $g_{3,k} \sim t^\gamma$. The 
value of $\gamma$ decreases from $\approx .81$ for $N=100$ to a value  close to $.66$ at $N=400$. An 
extrapolation gives the asymptotic value $\gamma=.6\pm.03$. Finally, and in analogy to the behaviour of $g_1$, we expect a 
crossover of $g_{3,k}$ to linear behaviour, since the knot eventually has to follow the whole polymer. 
This crossover has not happened yet for the times we were able to simulate.
In Fig. \ref{fig1} the vertical dashed line indicates the Rouse time, which we have estimated from the relation $g_3(\tau_R,N)=R_G^2(N)$. We checked that the Rouse time calculated in this way indeed grows as $N^{2.2}$. We thus see that, as was the case for the length autocorrelation, $\tau_R$ doesn't play a special role for $g_{3,k}$, and moreover, it seems that the second crossover occurs on a much slower time scale, which we expect to be $\tau_{T}$. In analogy with (\ref{4}), we therefore propose this second crossover to be of the form
\begin{eqnarray}
g_{3,k}(t,N) \simeq t^\gamma H(t/\tau_{T}) \hspace{1cm}Ê t > \tau_1
\label{6}
\end{eqnarray}
where $H(x)$ becomes constant for small $x$.
Since for $t > \tau_{T}$, the behaviour of (\ref{6}) must cross over into that of (\ref{1}), 
we obtain the relation $(1-\gamma)z_{T}=1$. Using the estimate $z_{T}=2.32\pm.1$, this leads to $\gamma=.57\pm.02$ consistent with our estimate. In Fig. \ref{fig2}, we show a 
sca\-ling plot of our data for $g_{3,k}(t,N)$ for various $N$-values (leaving out the initial power law 
regime) and using the values $z_{T}=2.35, \gamma=.57$. The scaling is rather well satisfied.
So, the available numerical evidence is consistent with the identification of the second crossover time with $\tau_{T}$.


In conclusion, by the first calculation in which dynamical information on 
the knot itself was monitored, we provided convincing evidence, that the presence of a 
knot in a ring polymer introduces a new slow topological time scale that is due to a self-reptation of 
the knotted region. 
Our results thus show the physical importance of the fact that the length of a knot in good solvent is a scale invariant, fluctuating quantity.

The picture presented here  implies that for a polymer below the $\theta$-point, where the knot delocalises ($\sigma=1$) \cite{BEAF05}, the topological time becomes proportional to the reptation time of the whole globule $\tau_d \sim N^3$. Moreover, in cases of strong localization, as for flat knots, we expect that the extra, slow time scale should not be present. The verification of these predictions and the experimental search for the topological time scale in specific processes are challenges for future work.

This work was supported by FIRB01, MIUR-PRIN05 and the FWO-Vlaanderen.







\end{document}